\documentclass{emulateapj}
\usepackage{graphicx}

\slugcomment{Accepted by AJ June 24 2008}

\begin{document}

\title{Radio Detection of Radio-Quiet Galaxies}

\author{J. A. Hodge, R. H. Becker\altaffilmark{1}}
\affil{University of California, 1 Shields Ave, Davis, CA 95616}
\email{hodge@physics.ucdavis.edu}

\author{R. L. White}
\affil{Space Telescope Science Institute, 3700 San Martin Drive, Baltimore, MD 21218}

\author{W. H. de Vries}
\affil{Lawrence Livermore National Laboratory, L-413, Livermore, CA 94550}

\altaffiltext{1}{Lawrence Livermore National Laboratory, L-413, Livermore, CA 94550}

\begin{abstract}
We investigate the radio emission of $\sim$ 185,000 quiescent (optically unclassifiable) galaxies selected from the Sloan Digital Sky Survey (SDSS).  By median-stacking FIRST cutouts centered on the optically-selected sources, we are able to reach flux densities down to the 10s of $\mu$Jy.  The quiescent galaxy sample is composed of two subgroups inhabiting vastly different regimes: those targeted for the SDSS MAIN Galaxy Sample ($\sim$55\%), and those targeted for the Luminous Red Galaxy (LRG) sample ($\sim$45\%).  To investigate the star-formation of these quiescent galaxies, we calibrate a radio-SFR conversion using a third sample of star-forming galaxies.  We confirm a tight power-law dependence for the star-forming (SF) sample, where L$_{1.4 GHz}$ $\sim$ (SFR)$^{1.37}$.  Comparing this SFR-indicator with indicators in the optical and UV, we derive conflicting SFR estimates for the MAIN sample quiescent galaxies.  These radio-derived SFRs intersect those calculated using the 4000-\AA\ break (D4000) around an SFR of 1 M$_{\sun}$ yr$^{-1}$  and agree to within a factor of 3 over the range of SFRs.  However, we find that the radio-derived SFRs are too high relative to the SFRs estimated for similar populations of galaxies using analysis of UV emission, implying either contamination of the radio by Active Galactic Nuclei (AGN) or incomplete dust modeling.  If AGN activity is dominant in these galaxies, then a relation between AGN radio luminosity and galaxy mass is required to explain the observed trends.  For the LRGs, on the other hand, we find the radio luminosity to be relatively high (compared to the SF galaxies) and independent of SFR as derived from D4000, indicating an AGN component dominates their radio emission.  AGN-based radio emission often implies the existence of radio jets, providing evidence of a mechanism for low-level feedback in these quiescent LRGs.  
\end{abstract}

\section{Introduction}

For normal galaxies -- those not deriving their radio luminosity from a nuclear ``monster" -- the radio emission is a probe of recent star-formation \citep{con92}.  In particular, the emission below 30 GHz is thought to be largely dominated by synchrotron radiation from ultrarelativistic electrons accelerated in supernova remnants.  As most of the remnants responsible for this non-thermal emission result from massive, short-lived stars, the resulting radio emission tracks recent star-formation.  Further evidence in favor of this claim comes from the existence of a tight radio/far-infrared (FIR) correlation, which has been found to hold over several orders of magnitude in radio luminosity \citep{har75, cram98, yun01}. 

The complex astrophysics connecting the integrated radio emission to supernova explosions has led to a historical preference for other methods.  Normal galaxies, with quite low radio flux densities, are also usually below the sensitivity of typical wide-area radio surveys.  Alternative indicators of star-formation rate (SFR) include continuum emission in both the near-ultraviolet and the far-infrared \citep{don84, har73}.  Another common diagnostic, referred to later in this paper, is the use of integrated emission-line fluxes \citep{coh76, ken83}.  (For additional indicators, see the excellent review by \citet{ken98}.)  

However, star-formation is not the only process that produces radio emission; Active Galactic Nuclei (AGN) can also be strong emitters in the radio.  This ambiguity has led to much debate, particularly for the less well-studied sub-mJy population.  On the one hand, there is a widely held belief that the population of sub-mJy radio sources is dominated by star-forming galaxies \citep{con84, win85, kro85, wal97}. For example, using an infrared-selected sample, \citet{boy07} found the radio/FIR correlation persisted down to $\mu$Jy flux densities, and they conclude that the radio emission of these sources is dominated by star-formation  (with redshift still unknown).  However, the hypothesis that there are a significant number of AGN at these low flux densities has recently received some attention as well.  \citet{jar04}, in a theoretical paper, suggested that the flattening in the observed counts of the normalized local luminosity function below S $\sim$ 1 mJy is due to a ``radio-quiet" AGN population.  In a follow-up observational study, \citet{sim06} concluded that radio-quiet AGN comprise a significant population below $\sim$ 300 $\mu$Jy, and that they may, in fact, dominate.  

Studying the general properties of the low-luminosity radio population, however, has proven to be difficult; the perennial trade-off between coverage area and depth discourages surveys that are both deep and wide.  In this paper, we take advantage of the overlapping coverage of two surveys, FIRST (Faint Images of the Radio Sky at Twenty Centimeters) and the SDSS (Sloan Digital Sky Survey), to study the faint radio emission of a large sample of optically-normal (``quiescent") galaxies.  By selecting galaxies with no strong optical emission lines from star-formation or AGN activity, we construct a sample that is likely to be inherently radio-quiet in nature.  Any radio emission we do find, then, must result from either extremely low-level SF and/or AGN activity, or activity that is obscured in the optical. By observing how this emission behaves as a function of other SFR indicators, we will attempt to discern the origin (SF or AGN) of the radio flux, and, in the absence of AGN activity, the radio emission will give us a handle on the amount of star-formation occurring in these galaxies.  

\citet{sal07} found that galaxies with no H$\alpha$ detectable in their SDSS spectra were mostly normal early-type galaxies with predominantly red optical colors, though some have relatively blue UV to optical colors (see also: Rich et al. 2005).  They calculated that some of these galaxies had UV-derived SFRs as high as 1 M$_{\sun}$ yr$^{-1}$, and a significant portion of those showed structure, like traces of faint spiral arms, or disturbed light profiles.  They identified an evolutionary sequence from star-formation to quiescence via nuclear activity.   Using a sample of nearby, early-type galaxies in the MOSES project (Morphologically Selected Ellipticals in SDSS), \citet{schaw07} reached a similar conclusion.  \citet{yan06} looked at optically-quiescent galaxies with red rest-frame colors and found that they were virtually indistinguishable in the color-magnitude-concentration space from galaxies classified as low-ionization nuclear emission-line regions (LINERs).  Still, very little has been done on the radio properties of optically-quiescent galaxies, and we intend to use this relatively unexplored wavelength regime to add to our knowledge of such galaxies.  

For the analysis of such faint radio sources, we have utilized the stacking technique presented in \citet{whi07}, hereafter referred to as Paper I.  This technique significantly reduces the noise in the combined images, allowing the investigation of median radio properties for sources significantly below the FIRST detection threshold without requiring countless more hours of telescope time.  In addition, stacking all of the fields guarantees that the radio data will not be restricted by survey limits.  In de Vries et al. (2007; hereafter, Paper II) we applied this method to the analysis of low-luminosity AGN, concluding that radio emission from star-formation can dominate over that due to the AGN emission in composite systems.

We begin by  describing the radio data and the details of the sample selection (\S \ref{data}).  We then go into the details of the stacking method (\S \ref{stack}).  In \S \ref{results} we present our analysis and a detailed discussion of the results; \S \ref{conclusions} contains our concluding remarks.  Throughout this paper we use the standard FRW cosmology of H$_0$ = 70 km s$^{-1}$ Mpc$^{-1}$,
$\Omega_{\Lambda}$ = 0.7, and $\Omega_{M}$ = 0.3.

\section{Sample Construction}
\label{data}
\subsection{The Radio Data}

The 1.4 GHz radio data we present are taken from the VLA FIRST Survey \citep{beck95}, which includes 9,030 $\deg^2$ sky coverage of the North Galactic Cap and Equatorial Strip, making it ideal for comparison with targets from the Sloan Digital Sky Survey (SDSS).  With only $\sim$ 90 sources $\deg^{-2}$ at the $5\sigma$ detection limit, 99.98\% of the 5 billion beam areas are `blank' (Paper I). The data, obtained from 180 sec VLA snapshots, have a typical rms of 0.15 mJy beam$^{-1}$.  By stacking 3000 such snapshots, we can reduce the noise in the resultant image to approximately 3 $\mu$Jy beam$^{-1}$ and achieve the sensitivity equivalent of 150 observing hours.  To convert to rest-frame 1.4 GHz luminosity, we perform a k-correction assuming an average radio spectral index $\alpha$ = -0.5.  This index provides a compromise between steep-spectrum radio-lobe emission and flatter spectrum emission from the AGN cores or star-forming regions, and it is the same index adopted throughout Paper II.

\subsection{The Quiescent Galaxy Sample}

The sample of quiescent galaxies was drawn from the SFR Catalog of the MPA/JHU VACs \citep{brinch04}.  The SFR catalog is a super-set of the galaxies discussed in \citet{brinch04} and contains all galaxies in the SDSS DR4 spectroscopic survey.  DR4 of the SDSS has 6851 $\deg^2$ of imaging and 4681 $\deg^2$ of spectroscopic coverage.  (See \citet{ade06} for more information on DR4).  Our final sample of quiescent galaxies is complete to a stellar mass lower limit of 10$^8$ M$_{\sun}$.  To accurately classify the objects, Brinchmann et al. reanalyzed the 1-D SDSS spectra using their own optimized pipeline as outlined in \citet{tre04}.  A galaxy was classified as either star-forming (SF), low S/N SF, AGN, low S/N AGN, Composite (C), or Unclassifiable based on the line ratios of the four emission lines described in Baldwin, Phillips, \& Terlevich (1981, hereafter BPT).  Composite galaxies have line ratios intermediate between SF and AGN.  For the SF and low S/N SF classes, the fiber SFR was obtained from detailed modeling of the emission lines, while for the AGN, C, and Unclassifiable galaxies, the fiber SFR was estimated by calibrating the relation between the specific SFR (calculated for the star-forming galaxies) and the measured D4000 value as defined in \citet{bal99}.  We will refer to the latter as ``D4000" SFRs to differentiate them from SFRs obtained via emission lines. For both types, total SFRs were calculated using an empirically based aperture correction scheme based on the likelihood distribution of the specific star formation rate for a given set of colors \citep{brinch04}.

To single out only ``quiescent" galaxies,  we kept only those galaxies catalogued as Unclassifiable. The majority of this class are galaxies with no or very weak emission lines \citep{brinch04}.  As the SDSS uses fiber-fed spectrographs, this is a `nuclear' classification.  The resultant sample contains $\sim$ 185,000 quiescent galaxies.  

\clearpage

We obtain additional parameters of interest for each galaxy from the NYU Value-Added Galaxy Catalog (NYU-VAGC).  As presented by \citet{bla05}, the NYU-VAGC is a collection of galaxy catalogs derived from the SDSS.  The latest version of the catalog accompanies the SDSS DR4.  We also make use of the derived ugriz absolute Petrosian magnitudes available in the associated catalog of K-corrections \citep{bla07}.  As these data were originally computed using a different value for the Hubble constant (H$_0$ = 100), we adjusted the derived absolute magnitudes to agree with the standard cosmology used elsewhere in this paper. 

\subsection{The Star-forming Galaxy and AGN Samples}
 
The star-forming sample, first introduced in Paper II, consists of all galaxies classified as star-forming (SF) in the DR2 SFR catalog.  This sample has S/N $>$ 3 in all four emission lines of the BPT diagram and falls below the conservative threshold for AGN rejection.   Star-formation rates quoted here were derived by \citet{brinch04} via careful modeling of the emission lines, and they take into account dust attenuation and fiber-corrections.  Note that these SFRs assume a Kroupa IMF \citep{kroupa01}.  For absolute magnitudes and derived parameters, we consulted the Stellar Mass Catalog as presented by \citet{kau03b}. 

The AGN sample was taken directly from Paper II and consists of objects from the \citet{kau03a} AGN compilation.  It should be noted that they use a more liberal AGN cut than \citet{brinch04}, and this sample therefore consists of objects classified as AGN as well as those classified as Composite in Binchmann et al.  This is a simple naming-convention issue and does not affect our results.  It is important to remember, however, that objects in the AGN sample are affected by star-formation to different degrees, with those closest to the cut having emission-line ratios more commonly associated with HII regions.  We use the D4000 SFRs (as we do with the quiescent galaxies) from \citet{brinch04}.

\begin{figure}
\centering
\includegraphics[scale=0.3]{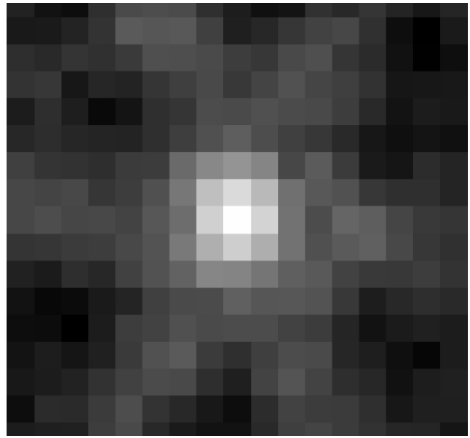}
\caption{A median image resulting from stacking 3000 quiescent LRGs.  The image is 0.5' $\times$ 0.5', (larger than normal, so the reader may observe the obvious sidelobes), and each pixel is 1.8" on a side.  The cutouts comprising this image correspond to the galaxies making up the right-most circle in Figure \ref{fig:sfr4}, prior to being converted to radio luminosity.  The image has a peak flux density of 90 $\mu$Jy, with an rms of 6 $\mu$Jy beam$^{-1}$.  This image is virtually indistinguishable from any of the other high S/N images generated for this study, independent of the type of galaxy.}
\label{fig:fitsimages}
\end{figure}

\section{Stacking FIRST images}
\label{stack}

Using the sample selection criteria described above, we compiled our optically-selected quiescent galaxy sample.  We then obtained 1.4 GHz data centered on each target position from the FIRST survey database.  These square ``cutouts" of the radio sky were 0.2' on a side and had the typical FIRST rms of $S_{rms}$ = 0.15 mJy beam$^{-1}$; these cutouts constitute our 3-D ``image block".  The size of the images was chosen pragmatically based on limitations of IDL memory.  Since all the analysis is based on the value of the central pixel, the image size is not important.  In any case, all stacked images have a FWHM $\le$ 8" and hence easily fit within the image format.  These blocks can be sorted by a variety of parameters (e.g., redshift, color, SFR, etc.) and stacked to yield median radio flux density/luminosity as a function of the parameter of interest, with a bin size ($N_{obj}$ / bin) determined by the radio survey depth required ($S_{min}$ $\sim$ $S_{rms}$ / $\sqrt{N}$).

Bins typically contained either 1500 or 3000 images each.  The median of the parameter of interest was determined for each bin.  To obtain the associated radio luminosity, the cutouts were processed through an IDL procedure which converted each individual image to units of radio luminosity using the target galaxy's measured redshift (Paper I) and then output the median image and $\pm 1 \sigma$ upper/lower bounds for each pixel. The basic algorithm for determining the error bars involved computing the probability that the true median lay between elements (k,k+1) in the sorted image stack using binomial probabilities as described in \citet{gott01}.   As these sources are all quite faint in the radio, we will simply quote the radio luminosity derived from the central pixel in each co-added image as done in Paper I.  

The median was used as opposed to the mean because it is more statistically robust in the presence of outliers and/or contamination.  For example, \citet{sal07} note that $\sim$ 2\% of the unclassifiable galaxies have line emission, some quite significant.  The median will be little changed by a 2\% contaminant.  The mean, on the other hand, is sensitive not only to outliers, but also to the threshold values imposed to try to weed out those outliers (Paper I).  When we used means, we found the radio luminosities all increased by 1-2 orders of magnitude.  That being said, the interpretation of the median is slightly more complicated.  The median shifts toward the `local' mean in the case of low S/N data.  (See Paper I for a more detailed discussion).  

Once the radio data were stacked, we corrected for the so-called `snapshot bias'.  This phenomenon is somewhat analogous to the `CLEAN bias' reported for the FIRST and NVSS surveys, where the result of the non-linear CLEAN algorithm is that a small amount of flux from the above-threshold sources is redistributed over the field \citep{beck95, con98}.  A similar bias for sub-threshold sources (that have therefore never been CLEANed) was discovered and calibrated in Paper I.  For sources fainter than 0.75 mJy, only 71\% of the true mean flux density is recovered per bin, while sources above 0.75 mJy are well-described by the addition of a constant 0.25 mJy CLEAN bias.  Since the majority of the target galaxies in this study are sub-threshold, we accounted for this snapshot bias via the constant multiplicative factor of 1.40.  

An example of a median-stacked image is shown in Figure \ref{fig:fitsimages}.  The image is 0.5' on a side and resulted from stacking 3000 quiescent LRGs.  The cutouts that went into this image were the flux density maps of the galaxies that were later converted to units of luminosity and stacked to give the right-most quiescent LRG bin (circle) in Figure \ref{fig:sfr4}.  The image peaks at  90 $\mu$Jy and has an rms of 6 $\mu$Jy beam$^{-1}$. 

\begin{figure}
\centering
\includegraphics[scale=0.45]{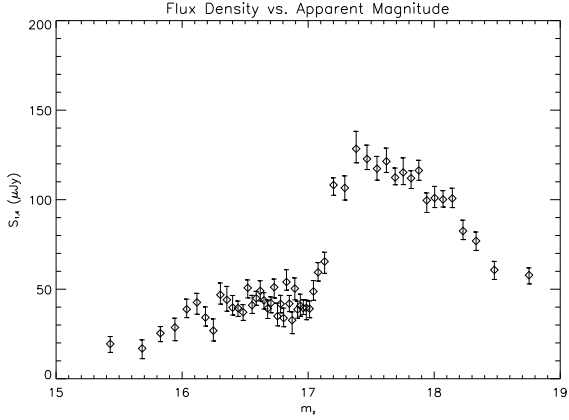}
\caption{A plot of radio flux density versus apparent magnitude for the entire quiescent galaxy sample.  The median-stacking method allows us to probe flux density levels down to the 10s of $\mu$Jy.  The clear selection bias toward stronger radio sources is evident below $m_z$ $\sim$ 17; this is a result of the SDSS target selection algorithm (see text).}
\label{fig:flux}
\end{figure}

\begin{figure}
\centering
\includegraphics[scale=0.45]{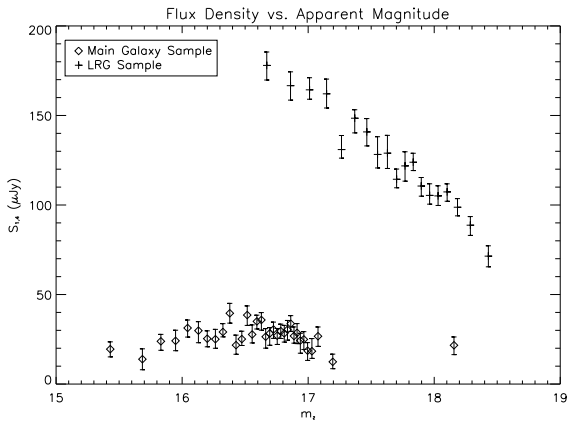}
\caption{A plot of radio flux density versus apparent magnitude for the quiescent galaxy sample, this time with galaxies targeted as part of the MAIN Galaxy Sample plotted separately from those targeted as LRGs.  The LRGs have much higher median flux densities.}
\label{fig:flux_select}
\end{figure}

\section{Results}
\label{results}

\subsection{Quiescent Galaxy Sample Selection Effects}

In order to explore the radio emission of the quiescent galaxy sample, we used the median-stacking technique detailed in \S \ref{stack}. Typical radio flux density levels attainable via this method are shown in the plot of flux density vs. apparent magnitude (z-band) in Figure \ref{fig:flux}.  Each point is the result of median-stacking 3000 quiescent galaxies.  As Figure \ref{fig:flux} demonstrates, a pronounced selection effect  becomes apparent when the data are represented in this manner.  This effect is the result of the SDSS target selection algorithm which includes different samples; the MAIN Galaxy Sample, the Luminous Red Galaxy (LRG) sample, and the Quasar, Star, or Serendipity samples \citep{sto02}.  Galaxy targets are selected for inclusion in the SDSS MAIN galaxy sample if they satisfy a number of criteria outlined in detail in \citet{str02}.  Approximately 55\% of our sample of quiescent galaxies were targeted as part of the MAIN Galaxy sample.  The remaining 45\% of our quiescent sample, save a small percentage of miscellaneous objects which we discarded, were targeted as part of the LRG sample.  As described by \citet{eis01}, the LRG sample extends to higher z and to fainter magnitudes than the MAIN galaxy sample and consists of objects that are luminous (L $>$ 3L$_*$) and intrinsically very red.   

Objects may also have more than one target flag set at a time.  Of particular concern to us are those objects targeted for both the MAIN and LRG samples.  Since the LRG group is generally more restrictive, and for reasons that will become apparent in our \S \ref{results}, we have chosen to keep these objects with the LRGs rather than the MAIN galaxy sample.  As advised in \citet{eis01}, we then keep only those objects with the GALAXY\_RED flag set and redshifts z $>$ 0.2.  These further restrictions account for the breakdown of the luminosity cut at lower redshifts and ensure the purity of the LRG sample.

If the data are first segregated into MAIN galaxies and LRGs, a rather different picture emerges.  Figure \ref{fig:flux_select} again shows flux density vs. z-band apparent magnitude, but this time the sources targeted as part of the MAIN Galaxy Sample are plotted as diamonds, while those targeted as part of the LRG sample are plotted as crosses.  The MAIN Galaxy Sample show typical radio flux densities in the 10s of $\mu$Jy, with an increasing number of Southern Equatorial Strip sources comprising the low-magnitude tail. Of these galaxies, fewer than 0.1\% were found to be coincident with cataloged FIRST sources.   The LRGs have significantly higher median radio flux density levels (though still far below the FIRST threshold for detection) and exhibit a trend toward lower flux densities with increasing magnitude.  Out of the LRG sample, 0.9\% are found to be within 2" of a source in the FIRST catalog, a detection rate nearly ten times that of the MAIN Galaxy Sample.

\begin{figure}
\centering
\includegraphics[scale=0.45]{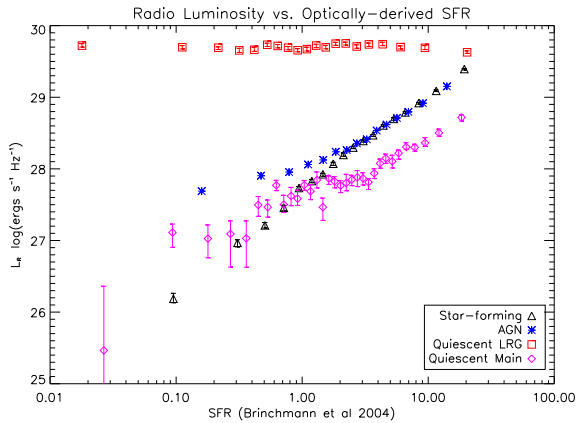}
\caption{Radio Luminosity (1.4 GHz) vs. optically-derived SFR from \citet{brinch04}.  This plot shows the star-forming galaxies and AGN originally presented in Paper II, as well as the quiescent MAIN sample galaxies and LRGs overplotted.  Note that all of these objects are still at least 3 orders of magnitude fainter than the typical radio-loud object.}
\label{fig:lumvsSF}
\end{figure}

\subsection{Radio Luminosity as a Proxy for Star-Formation}

As mentioned in the Introduction, radio emission can be used as a tracer of star-formation for galaxies with no contaminating nuclear activity.  We use the star-forming galaxy sample, described in Section \ref{data}, to obtain the relationship between star-formation rate and radio luminosity.  The SF galaxies have measured SFRs based on H$\alpha$ and other optical emission lines,  and we sorted the sample by this ``H$\alpha$-derived" SFR and binned/stacked the data using the method outlined in Section \ref{stack}.  The result, published in Paper II and repeated in Figure \ref{fig:lumvsSF}, shows a tight correlation between radio luminosity and SFR for the SF galaxies which extends over more than two orders of magnitude in SFR.  The least-squares fit to the data gives log(L$_{1.4}$) = (1.37 $\pm$ 0.02) log(SFR) + (27.67 $\pm$ 0.01). 

There have been a number of studies deriving SFR from 1.4 GHz emission, many of which use the radio-SFR relation presented by \citet{con92}.  This relation, and those like it \citep{yun01} assume a linear relationship between SFR and radio luminosity.  Indeed, the tightness of the radio-FIR correlation was first interpreted as evidence for a linear relationship.  However, several authors \citep{Pri92,Bell03} have shown that the synchrotron component, which generates $\sim$ 90\% of the emission at 1.4 GHz, depends non-linearly on the SFR.  \citet{Pri92} found that, by decomposing the radio emission in the radio-FIR relation into thermal bremsstrahlung and non-thermal synchrotron components, the non-thermal radio luminosity scaled as $(SFR)^{1.2}$, even while the thermal emission was directly proportional to SFR.  \citet{Bell03} argued that the radio-FIR correlation is a conspiracy resulting from the non-linearity of both SFR indicators.  They find that $L_{1.4}$ goes as $(SFR)^{1.3}$, a result which is very close to our derived power-law dependance of 1.37.

\subsection{Radio Emission from Quiescent Galaxies}

Figure \ref{fig:lumvsSF} shows our results for the quiescent galaxies overplotted on the SF galaxies and AGN of Paper II.  The first thing to notice is the large difference between the behavior of the MAIN and LRG-targeted samples.  Keeping in mind that the LRG bins do have significantly higher median redshifts than the MAIN galaxy bins, the samples nevertheless exhibit very different trends.  The LRG sample maintains relatively constant median radio luminosity even with optically derived SFRs (from D4000) varying from 0.1 to 10 M$_{\sun}$ yr$^{-1}$, suggesting that the radio emission from these optically-normal LRGs is dominated by AGN activity rather than star-formation.  I.e., if some of the radio luminosity came from star-formation, there would be a dependance on SFR.  Indeed, for LRGs with SFR $<$ 1 M$_{\sun}$ yr$^{-1}$, less than 1\% of the radio flux should originate from SF activity.  It has been pointed out to us that perhaps the SFRs in the MPA/JHU catalog were never intended to be appropriate for the LRGs in the catalog, seeing as they derived from star-forming galaxy colors and reach quite spectacular rates for galaxies believed to be red and dead.  This does not change our result.  Whether the LRG SFRs are in the ballpark or complete nonsense, the median radio luminosity of the sample is simply too high to be attributable to SF alone.

One may ask how LRGs actually classified as AGN compare.  We therefore compiled a sample of galaxies classified as AGN or Composite from the MPA/JHU SFR catalog that were also targeted as LRGs.  We found only $\sim$ 600 such galaxies, not even enough for a full bin and not nearly enough to examine trends in the stacked behavior.  The partial bin we were able to make lies basically right on top of the quiescent LRGs, with a median SFR of $\sim$ 4 M$_{\sun}$ yr$^{-1}$ and an L$_R$ of 10$^{29.9} $ergs s$^{-1}$ Hz$^{-1}$.

Figure \ref{fig:lumvsSF} therefore provides evidence that quiescent LRGs host AGN, a result that has some interesting implications.  Most current models of cosmic structure formation include AGN feedback, and both radiative and mechanical heating of the ``radio mode" AGN are now recognized as important in galaxy formation scenarios (for e.g., Croton et al., 2006, and Granato et al., 2004).  As radio emission from black holes often indicates the presence of jets, these data lend observational support to the existence of a mechanism for low-level feedback in these galaxies.  In addition, this result also gives a quantitative feel for the feedback subsequently implied.  The constant median value of L$_{R}$ $\sim$ 4 $\times$ 10$^{29}$ ergs s$^{-1}$ Hz$^{-1}$ radiated by these quiescent LRGs, while still technically low-luminosity, is over two orders of magnitude larger than that attributed to the ``ground state" of the optically-detected low-luminosity AGN in Paper II.  As to why the LRGs are so much more radio-luminous, perhaps this is evidence that more massive AGNs are more radio-luminous.  But speculation aside, it is true that here we are dealing with median statistics and therefore have no knowledge of the underlying luminosity distribution (i.e., whether it is bimodal, etc.).  We also cannot say whether the spread in the distribution is due to time variability or static differences between galaxies.  However, by treating this value of radio luminosity as a time-average (rather than ensemble-average), we get an estimation of the average luminosity of radio jets which may result in AGN feedback in LRGs.

\begin{figure}
\centering
\includegraphics[scale=0.45]{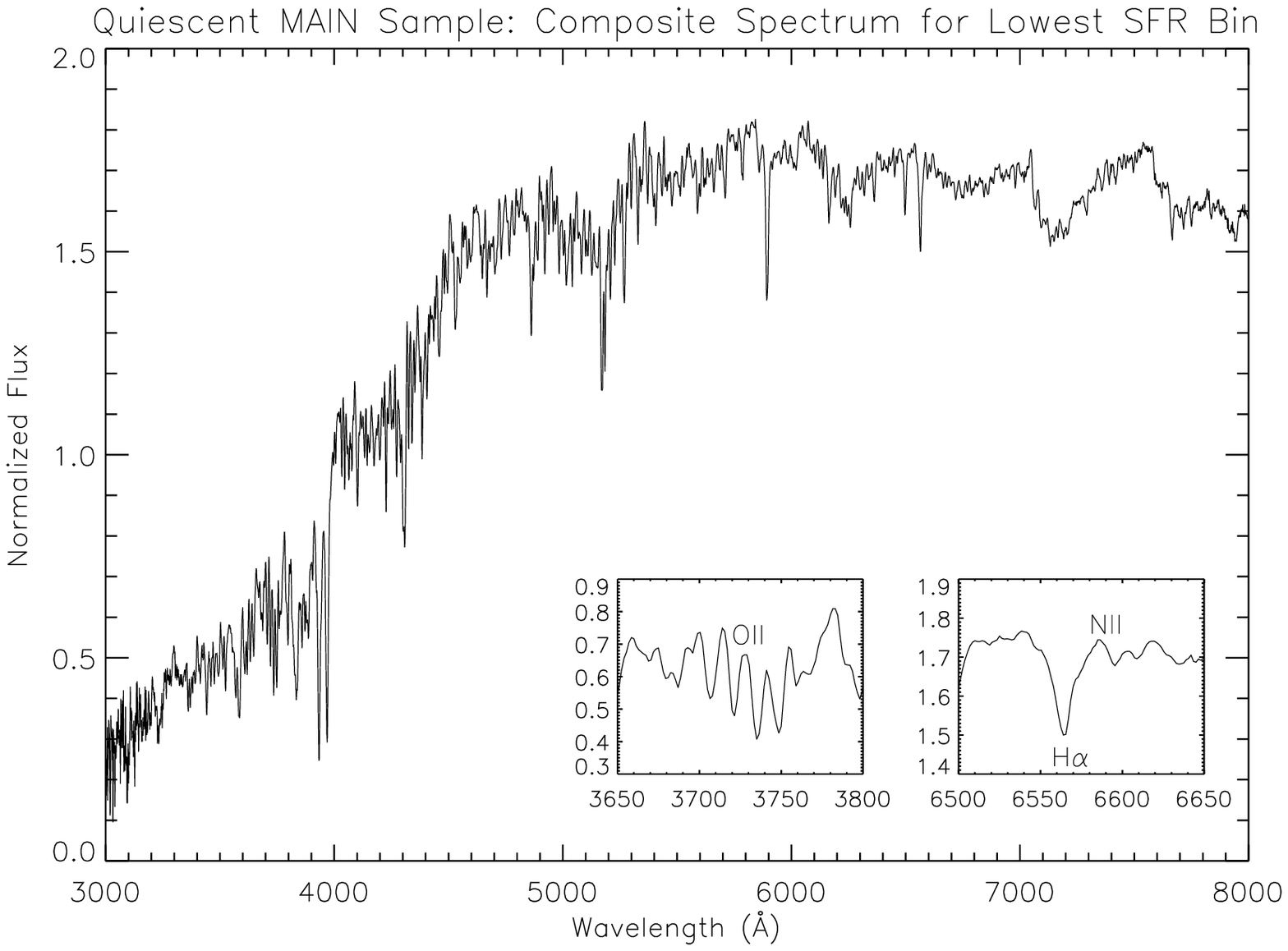}
\vspace{10pt}
\vfil
\includegraphics[scale=0.45]{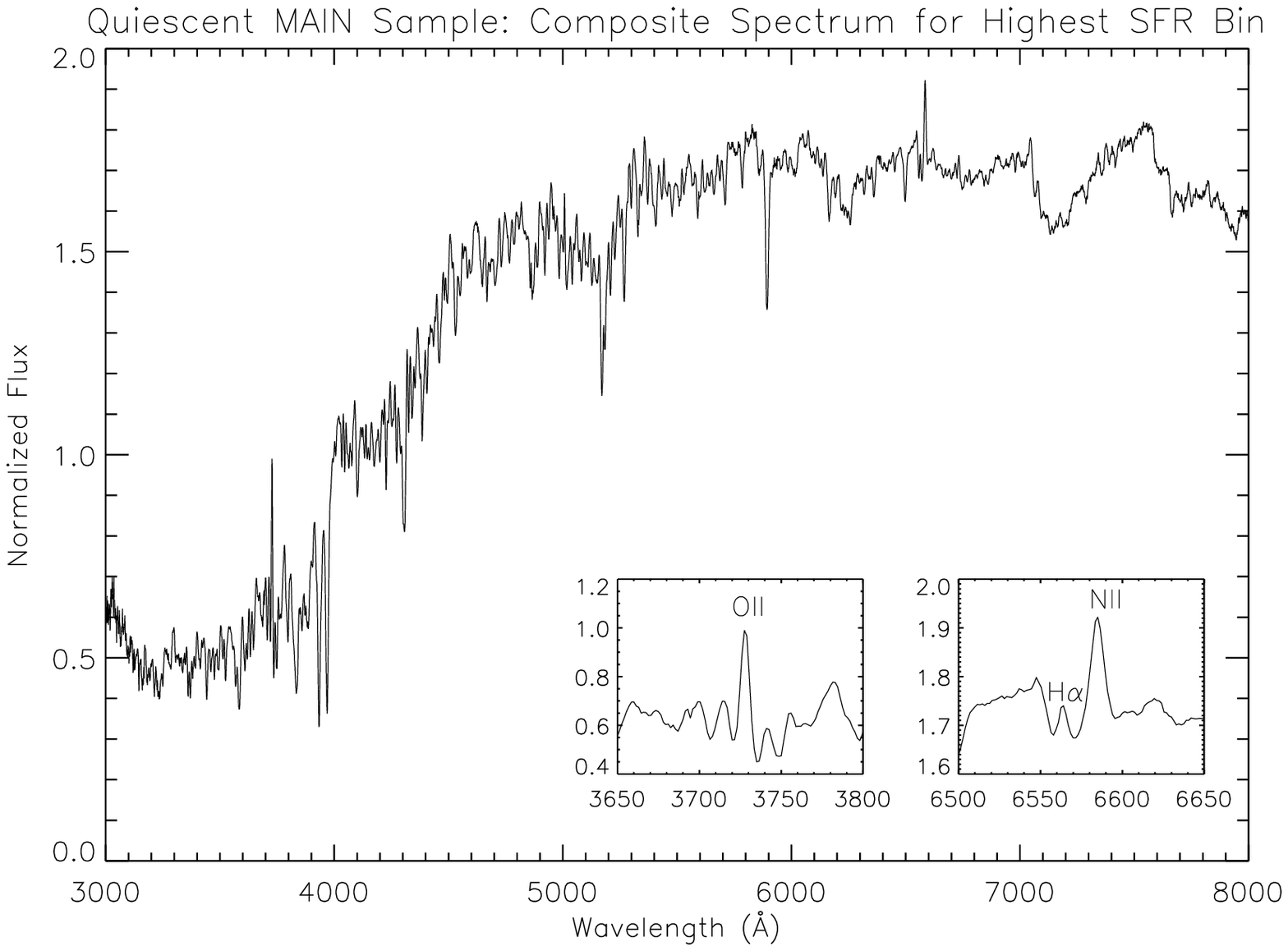}
\caption{Composite spectra for the MAIN sample quiescent galaxies.  The top panel (a) shows the composite spectrum for the 3000 galaxies comprising the bin with the lowest SFR in Figure \ref{fig:lumvsSF}, while the bottom panel (b) shows that of the bin with the highest SFR.  The insets show the spectral features associated with OII, $H\alpha$, and NII.}
\label{fig:spectra}
\end{figure}

We now turn our attention to the MAIN sample galaxies in Figure \ref{fig:lumvsSF}.  These galaxies show a trend in radio luminosity with SFR which intersects the SF sample around 1 M$_{\sun}$ yr$^{-1}$.  The agreement is worse for higher values of SFR, where the MAIN sample galaxies deviate systematically from the SF galaxies by a little over a factor of 3.  There are two competing theories for the origin of the radio emission from these galaxies:  that it is due to SF, or that it is due to AGN activity.  The fact that L$_{R}$ shows a dependence on D4000 SFR, and that it disagrees with the SF sample by no more than a factor of 3 over the whole range of SFRs, suggests that it may arise in part from star-formation.  Alternatively, one may argue we are seeing the results of a correlation between $L_R$ and mass that is due to AGN activity, a valid argument that we elaborate on below.  However, since the $L_R$ is near the level of the SF galaxies, we think it only fair to entertain the notion that it is due to SF.  If we assume this to be the case, the disagreement between the MAIN sample and SF galaxies may be an artifact of the different methods (D4000 vs. emission lines) used by \citet{brinch04} to measure SFRs for the different samples.  Although \citet{brinch04} proved the reliability of these D4000 values for galaxies in the star-forming regions of the BPT diagram, that doesn't guarantee that the calibration yields spot-on values for galaxies with suppressed emission lines.  \citet{sal07} argue that the D4000 - specific SFR calibration and the color-based aperture corrections are only completely accurate for the SF galaxies, since both methods are dependent on time-scale and other galaxy types have different star-formation histories. 

\begin{figure}
\centering
\includegraphics[scale=0.45]{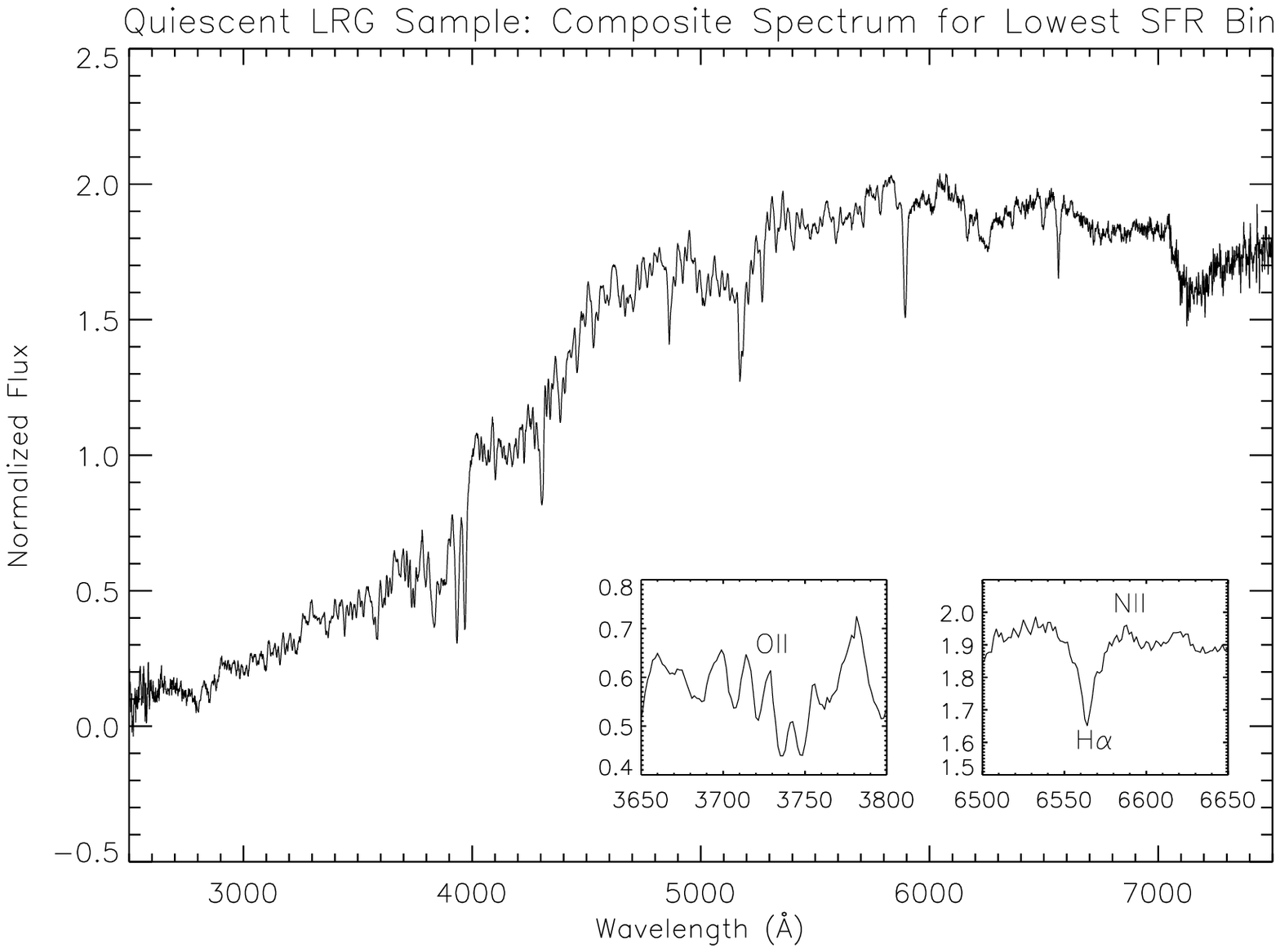}
\vspace{10pt}
\vfil
\includegraphics[scale=0.45]{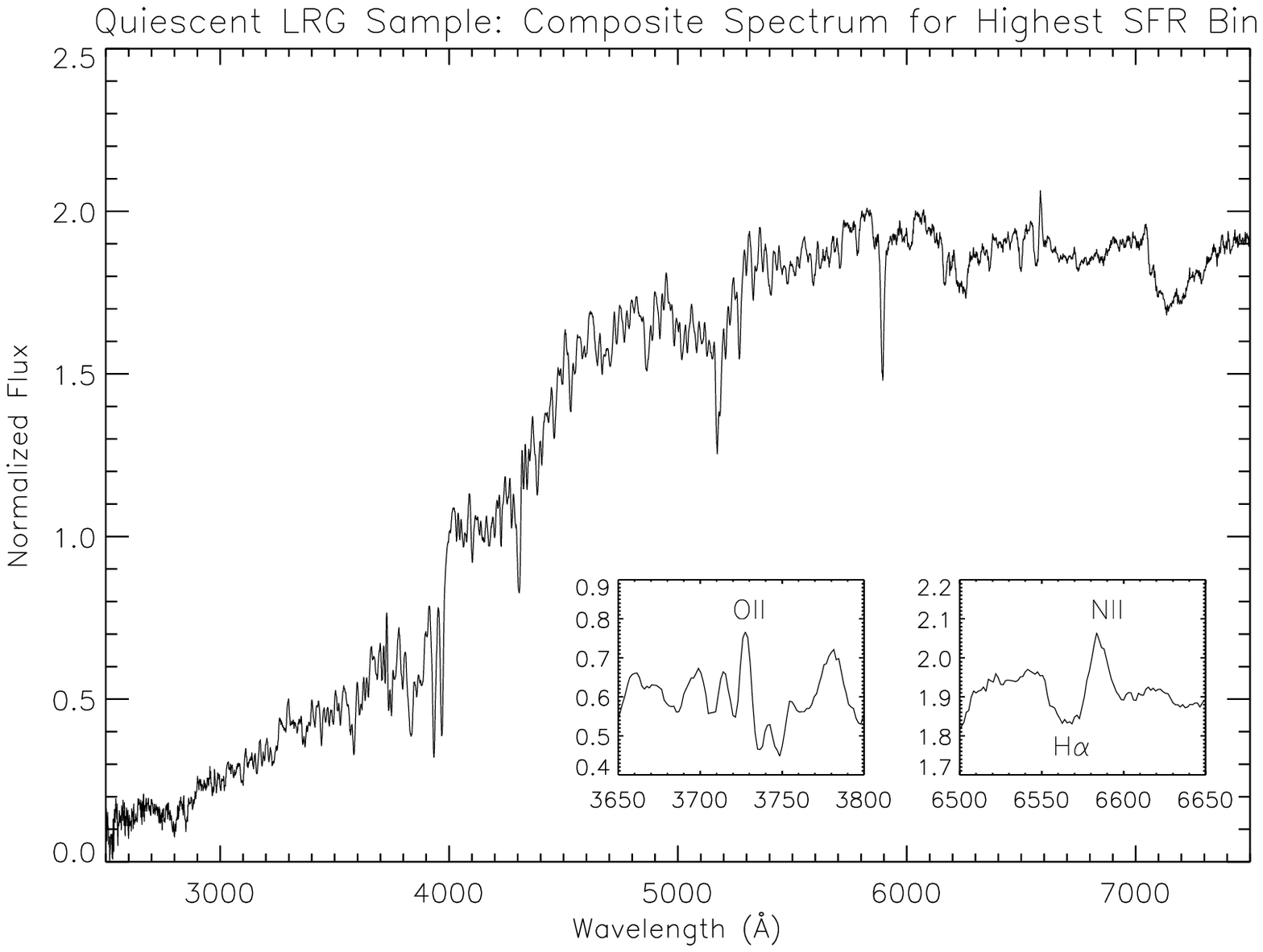}
\caption{Composite spectra for the quiescent LRG sample.  The top panel (a) shows the composite spectrum for all the galaxies comprising the bin with the lowest SFR in Figure \ref{fig:lumvsSF} as calculated from D4000, while the bottom panel (b) shows that of the bin with the highest SFR.  The insets show the spectral features associated with OII, $H\alpha$, and NII.}
\label{fig:LRGspectra}
\end{figure}

\begin{figure}
\centering
\includegraphics[scale=0.45]{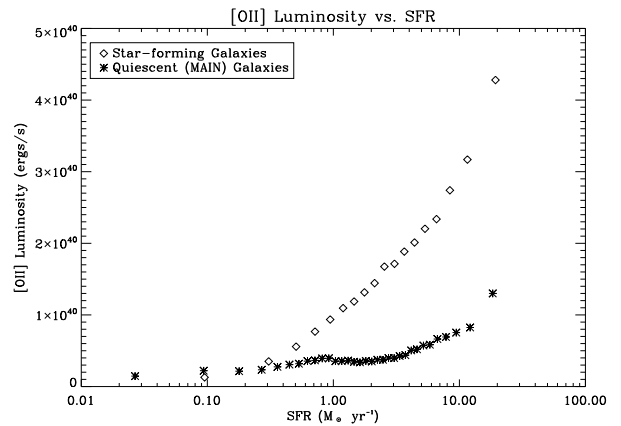}
\caption{[OII] Luminosity vs. SFR for the quiescent MAIN and star-forming galaxy samples.   Values for [OII] Luminosity are rough estimates using the conversion from equivalent width to luminosity in \citet{ken92} with median g'-band luminosity.}
\label{fig:oii_vs_sfr}
\end{figure}

The wary critic might ask whether D4000 is measuring star-formation at all in this case, or whether AGN emission is contaminating the D4000 break as well, leading to overestimation of the D4000 SFRs for the MAIN sample.  We have considerable reason to believe that this, at least, is not the case.  For one thing, the absence of observable line emission from the AGN makes it unlikely that the AGN would contribute significantly to the continuum near the 4000 \AA\ break.   And furthermore, even for optically-detected AGN, \citet{kau03a} showed that the AGN contribution to the blue continuum is ignorable.  

To get a better feel for these objects, we take advantage of our large sample size and examine trends in median-stacked spectra.  We have created median composite SDSS spectra for both the MAIN galaxy sample and the LRGs as a function of SFR bin as shown in \ref{fig:lumvsSF}.  These spectra, normalized such that the continuum value at 4000 \AA\ is unity, are shown in Figures \ref{fig:spectra} and \ref{fig:LRGspectra}, respectively.  In each case, the top panel shows the spectrum for the lowest SFR bin and the bottom shows that of the highest SFR bin.  We should note that we also tried binning spectra by SDSS z-band absolute magnitude, and the resulting composite spectra had line emission no more obvious than those spectra binned by SFR that we show here.

These high S/N composite spectra clearly show very weak emission features.  They do, however, show strong Balmer line absorption; strong enough to plausibly swallow most of the H$\alpha$ emission.  This is supported by some [NII] emission even in the absence of H$\alpha$.  The inset of Figure \ref{fig:spectra}b shows a zoomed-in view of this region for the high-SFR galaxies in the MAIN sample.  H$\alpha$ emission is just visible over the accompanying deep absorption feature.  Some may argue that the [NII]/H$\alpha$ ratio is more easily explained by a LINER (low ionization emission-line region) spectrum, and this may well be the case.  However, there is also an obvious turn-up below $\sim$ 3300 \AA\ in the spectrum of the high-SFR MAIN galaxies, and no sign whatsoever of a similar trend in the low-SFR MAIN galaxies or LRGs.

\begin{figure}
\centering
\includegraphics[scale=0.45]{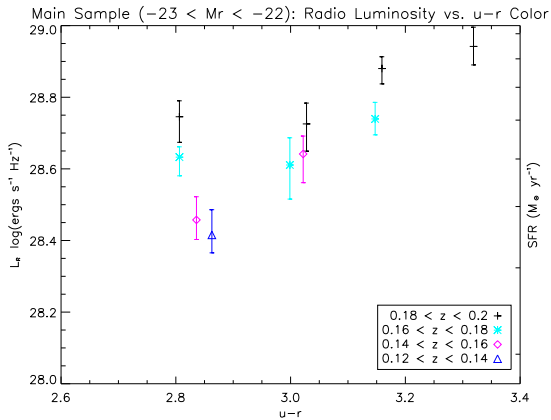}
\caption{Radio Luminosity (1.4 GHz) vs. u-r color for a volume-limited subsample of the quiescent MAIN galaxy sample.  Since the fraction of diluted AGN increases with redshift \citep{rev07}, the radio luminosity might be expected to increase if misclassified AGN dominate the radio emission.}
\label{fig:vollim}
\end{figure}

For both samples, the spectra with higher median SFRs show greater [OII] emission, a useful proxy for star-formation in the absence of H$\alpha$ \citep{ken92}.  (We should note, however, that \citet{yan06} found [OII] emission in red and post-starburst galaxies can be as strong as in star-forming galaxies.)  We have attempted to quantify this result for the MAIN sample galaxies in Figure \ref{fig:oii_vs_sfr}.  The [OII] Luminosities on the vertical axis were calculated by determining the equivalent widths and using the conversion to luminosity given in \citep{ken92}.  They are only rough estimates, however, as we used magnitude in the $g^{\prime}$-band rather than the B-band, and we used the median magnitude for the conversion of the composite equivalent width.  For comparison, we also plot the results for the Star-forming sample.  The [OII] Luminosities for the MAIN sample do increase with SFR, but only weakly relative to the Star-forming sample.  If we assume the [OII] in the quiescent MAIN sample traces SF just as reliably as in the SF sample, and if we use the SF galaxies to calibrate an [OII]-SFR relation, the implied SFRs are all $\le$ 1 M$_{\sun}$ yr$^{-1}$.  There also appears to be a kink around 1 M$_{\sun}$ yr$^{-1}$, the same point at which the radio luminosities diverge, prompting speculation of a change in population at that point.  Regardless, the weakness of the [OII] dependence on D4000 SFR may indicate that both SF and AGN activity are contributing to the line emission, with the AGN component dominating.  Of course, why these galaxies have such weak line emission has been the question all along, and it pertains to both AGN and SF.

\begin{figure}
\centering
\includegraphics[scale=0.45]{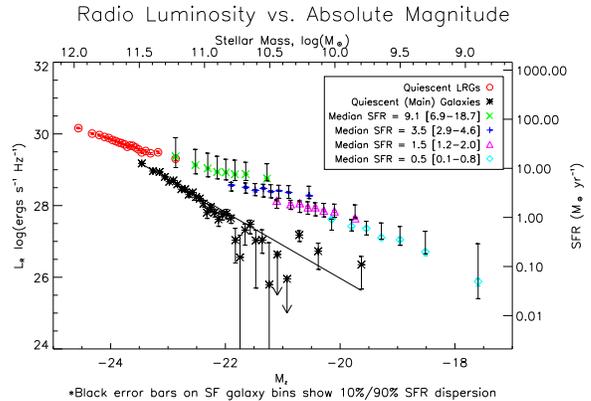}
\caption{Radio Luminosity (1.4 GHz) vs. z-band Absolute magnitude for quiescent and star-forming galaxies.  Arrows indicate upper limits.  The star-forming galaxies have been split into 4 bins based on their fiber-corrected star-formation rates, and the median SFRs and [SFR ranges] are shown in the legend.  We fit a line to the quiescent galaxy data; the least-squares fit gives log($L_R$) = (-0.92 $\pm$ 0.02)$M_z$ + (7.5 $\pm$ 0.4).  Additional error bars are overplotted on the star-forming galaxy points.  These larger error bars correspond to 10\% / 90\% of the SFR dispersion for the galaxies within each bin.}
\label{fig:sfr2}
\end{figure}

The real question here is which component (AGN or SF) dominates the radio emission, and we do not believe that these composite optical spectra alone can answer that question.  In a recent paper, \citet{mag07} use a sample of optically obscured, 24-$\mu$m-selected galaxies observed by Spitzer in the First Look Survey (FLS) to conclude that the 1.4 GHz luminosity is dominated by star-formation activity despite the presence of an AGN component.  IRAC photometry sets the majority (66\%) in a redshift interval of z = 1-3.  The same conclusion was drawn in Paper II for optically classified AGN/Composites in the local universe with SFR $>$ 1 M$_{\sun}$ yr$^{-1}$. 

If AGN activity is the source of the radio emission from the MAIN galaxy sample, then one possible explanation for the weak line emission is dilution.  It has been shown that stellar continuum light of a host galaxy may dilute the nuclear emission lines of an AGN and hinder its detection \citep{mor02}.  \citet{rev07} present evidence that this effect leads to an increasing number of misclassified weak-lined AGN with increasing redshift in the SDSS and other redshift surveys.  We test for this effect in Figure \ref{fig:vollim}, where we take a volume-limited subsample in a narrow optical luminosity range and assess the effect of redshift on radio luminosity versus u-r color.  Note that the magnitude cut of -22 $<$ $M_z$ $<$ -23 has preferentially selected the more luminous galaxies (in the radio as well as the optical).  If increasing redshift bins show larger luminosities, we might conclude that AGN are contributing significantly to the radio emission.  While there seems to be a slight trend, the effect is quite small and the error bars are large.  The null result is therefore inconclusive.

\begin{figure}
\centering
\includegraphics[scale=0.45]{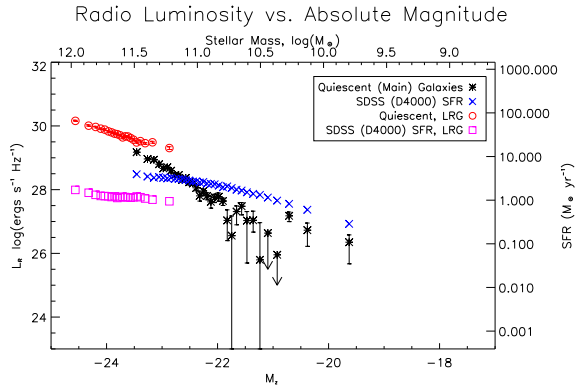}
\caption{Radio Luminosity (1.4 GHz) vs. z-band absolute magnitude for quiescent MAIN sample galaxies and LRGs.  SFRs from \citet{brinch04} are overplotted for each bin of both samples.}
\label{fig:sfr4}
\end{figure}

\begin{figure}
\centering
\includegraphics[scale=0.45]{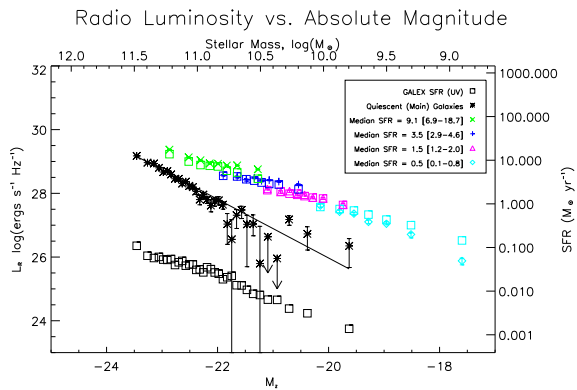}
\caption{Same as Figure \ref{fig:sfr2}, but with UV-derived SFRs from GALEX overplotted for each point as open squares.  Though the UV-derived SFRs and the D4000 SFRs are discrepant by 1-2 orders of magnitude for the MAIN galaxies, we find them to be tightly correlated, arguing for a common origin.}
\label{fig:sfr3}
\end{figure}

If the radio emission of the MAIN galaxy sample stems from star-formation, dilution might still be the culprit.  To test this theory, we plot radio luminosity versus z-band absolute magnitude for the quiescent and star-forming galaxy samples in Figure \ref{fig:sfr2}.  The SFR axis on the right-hand side of this figure (and those that follow) utilizes the radio-SFR calibration equation obtained in the previous section using the star-forming galaxy sample.  If we suppose for the moment that the radio emission of the Main sample quiescent galaxies originates with star-formation, then we can apply it to these galaxies as well (it obviously does not apply to the LRGs given our above results).  We derived the mass (upper) axis from a linear least-squares fit to the plot of log(stellar mass) versus $M_z$, where the masses came from the Stellar Mass Catalog \citep{kau03b}.  The star-forming galaxy sample was first sorted by SFR and split into 4 subgroups to aid in the interpretation of the quiescent galaxy SFRs.  The median SFRs and ranges in median SFR for the bins in a particular subgroup are shown in the legend.  The stacks were reduced to 1500 objects per bin to allow for an adequate number of data points in each subgroup, and the resultant subgroups are plotted as different symbols for easy differentiation.  The data for the LRG and MAIN samples were binned and stacked following our standard procedure, with the error bars representing $\pm 1\sigma$ in median radio luminosity.  We find that the quiescent MAIN sample follow a power-law fit of log(L$_{R}$) = (-0.92 $\pm$ 0.02)M$_{z}$ + (7.6 $\pm$ 0.5).  
 
While the H$\alpha$-based method does yield a large overall range of star-formation rates, the SF sample display a narrow range of SFRs for a given absolute magnitude of the host galaxy.  This result was previously confirmed by \citet{sal07}.  If we assume no AGN contamination for the moment, then using stacked radio luminosity as an indicator yields SFRs for the quiescent MAIN galaxies that are consistently 3-5 times lower than the SF galaxies.  Keep in mind that we are not suggesting that the quiescent MAIN sample galaxies are intrinsically the same as the line-emitting SF galaxies, which we already demonstrated to populate a tight sequence in SFR-M$_z$ space.  But if we are to believe the trend demonstrated in Figure \ref{fig:lumvsSF}, particularly for the low-L$_R$ end, then one interpretation is that the strength of H$\alpha$ needed for detection (and for subsequent SFR-estimation) is proportional to absolute magnitude.   This interpretation implies a sort of H$\alpha$-dilution.  To illustrate this point further, we have taken the star-forming galaxies in Figure \ref{fig:sfr2} and overplotted larger error bars representing the 10\% / 90\% range in SFR dispersion for the galaxies comprising each data point.  

Figure \ref{fig:sfr4} uses the same axes to show how the radio-derived star-formation rates for the quiescent galaxies compare to those available from \citet{brinch04} in a different way than Figure \ref{fig:lumvsSF}.  The 8-point stars still represent the quiescent galaxies targeted in the MAIN Galaxy Sample, while the objects in the LRG sample are shown as open circles in the upper left-hand region of the plot.  We have then overplotted (as crosses and squares, respectively) the median optically-derived SFR for each bin using the SFR axis on the right-hand side of the plot.  For the quiescent LRG sample, the D4000 SFRs are roughly two orders of magnitude lower than those inferred from the radio data; i.e., there is excess radio emission over that which is attributable to star-formation.  This is again consistent with the hypothesis that the radio emission from these LRGs is dominated by AGN activity.  

The data for the MAIN sample quiescent galaxies show somewhat better agreement.  The D4000 SFRs for the MAIN sample overlap the radio-derived SFRs at a few M$_{\sun}$ yr$^{-1}$ and an absolute magnitude of M$_{z}$ = -22.5, though the two estimates show different slopes with absolute magnitude.  If the quiescent MAIN sample are star-forming, this discrepancy could simply be explained by the different methods used to calculate SFR between SF and ``Unclassifiable" galaxies.  If the emission is dominated by an AGN contribution, alternatively, it must scale with galaxy stellar mass to explain the observed correlations.  This is not surprising for star formation in a homogenous sample of galaxies, but one might argue that this is expected for AGN as well, given the well-known scaling relation between black hole mass and bulge luminosity \citep{kor95, ferr00}.  Our own results for the quiescent LRGs may even support this claim.  

As we have discussed above, radio emission is an indirect indicator of star-formation.  The UV continuum, on the other hand, is thought to be directly proportional to the photospheric emission of young stars \citep{ken98}.  Still, it is not free from problems of its own, most notably significant dust obscuration.  Nevertheless, as another comparison, we overplot UV-derived SFRs from GALEX \citep{sal07} in Figure \ref{fig:sfr3}, corrected to our IMF.  As our samples do not completely overlap, we have simply taken the median of the UV SFRs available for each bin.  For our MAIN sample quiescent galaxies, there were generally $\sim$ 300 galaxies (per 3000-galaxy bin) with UV SFRs, and we find these galaxies to have fairly representative radio luminosities (i.e., there is no selection bias between our samples).  None of the quiescent galaxies targeted as LRGs had UV-SFRs available from this catalog.   

These UV-based SFRs were obtained by \citet{sal07} via SED-fitting of emission from the UV to the z-band with various models from the \citet{bruz03} population synthesis code.  Because the UV luminosity imposes the tightest constraint on the SFR, the authors refer to these SFRs as ``UV-derived".  Their quoted rates are averaged over a span of 100 Myr and constitute the first test of the two-component dust attenuation model proposed by \cite{char00}.  \citet{sal07} compare these rates with the optically-derived rates for the \citet{brinch04} star-forming sample and report good overall agreement, with some systematic deviation due to differences in modeling dust attenuation.  It is therefore not surprising that we also find good agreement in the star-forming sample, as they are derived from the same catalog and served as our radio-SFR calibrators.   

The interpretation of the UV and radio-derived SFRs for the MAIN sample quiescent galaxies is, once again, less straightforward.  The UV-derived SFRs are between 2-3 orders of magnitude lower than the SFRs derived from radio-stacking.  Given the rough agreement of the radio and D4000 SFRs, it is then no surprise that \citet{sal07} find a $\sim$ 2 order-of-magnitude discrepancy for the UV versus optically-derived SFRs of their ``No H$\alpha$" sample, which comprise roughly $\sim$ 10\% of our total MAIN quiescent sample.  They conclude that, since these galaxies fall largely on the red sequence, the UV SFRs are more realistic.  Another of their arguments is that such high SFRs as \citet{brinch04} estimate would result in a large fraction of their ``No H$\alpha$" galaxies having detectable H$\alpha$.  On the contrary, we have shown in Figure \ref{fig:sfr2} that this result could also be explained if the strength of H$\alpha$ required to constitute a ``detection" is simply a function of optical luminosity, with the levels predicted for these quiescent galaxies falling below the detection limit for a given absolute magnitude.  

The strongest argument for the UV-SFRs is that the 2-3 orders of magnitude discrepancy between the radio and UV would require an underestimation of dust extinction of 5-8 magnitudes.  This is, admittedly, quite large, but perhaps the disagreement between SFR estimates could be explained by the location of the star-formation;  If the star-formation is dominated by nuclear star-formation, it might be easier to hide both the line-emission and the UV flux from observation.  This is speculative, but we would like to emphasize again that the radio emission and D4000 index are two independent checks on the presence of star-formation, and they are consistent with one another to within a factor of $\sim$ 3.  While they both can be higher than the true SFR for completely different reasons, we feel this is a coincidence that requires further explanation before the possibility of SF in these galaxies can be thrown out altogether.  We will, however, freely acknowledge that if the GALEX SFRs are right, then for sure the radio-derived SFRs are due to AGN activity.  Overall, whether this discrepancy is ultimately the result of an AGN component to the radio emission or a poor dust-attenuation model is not a question we feel can be definitively answered here.

\section{Conclusions}
\label{conclusions}

We have investigated the radio emission from a sample of $\sim$ 185,000 optically-selected quiescent galaxies from the Sloan Digital Sky Survey.  By stacking FIRST cutouts centered on the galaxy positions, we reach median radio flux densities down to the 10s of $\mu$Jy.  We calibrate the relation between 1.4 GHz radio emission and optically-derived SFR using a sample of star-forming galaxies, allowing us to look for star-formation in their quiescent counterparts.  The main findings of this study are summarized below.

\begin{itemize}

\item We find a tight correlation between radio luminosity and SFR for the star-forming galaxy sample, with L$_{1.4GHz}$ going as (SFR)$^{1.37}$.  

\item The quiescent galaxy sample consists of two subgroups: those targeted as part of the MAIN Galaxy Sample, and those targeted as part of the Luminous Red Galaxy (LRG) Sample.  The former show partial overlap with the star-forming sample on a plot of L$_{R}$ vs. optically-derived SFR, while the latter show a relatively constant, high value of median radio luminosity that does not vary with SFR.

\item   For the LRGs, the correspondingly high rates of radio-derived star-formation conflict with those estimated by \citet{brinch04} via calibration of the 4000-\AA\ break by two orders of magnitude on a plot of derived-SFR vs. M$_z$.  This result, combined with the relative constancy of L$_{R}$ with D4000 SFR, suggests that the radio emission of these quiescent LRGs derives not from star-formation but from an AGN component.  This implies the possible presence of radio jets and a mechanism for low-level feedback in these galaxies.  

\item The MAIN sample galaxies, on the other hand, exhibit a L$_R$-SFR trend which intersects the SF sample around 1 M$_{\sun}$ yr$^{-1}$ and deviates by no more than a factor of 3 over the whole range of SFRs.  This agreement as well as the dependence of L$_{R}$ with SFR suggests that the  L$_{R}$ might arise in part from star-formation (particularly in so far as the AGN of Paper II show no such dependence.)  If we assume the radio emission of these galaxies originates from star-formation, the deviation at high/low values of SFR might then be due to the different diagnostic measures used to estimate SFR for the two samples (i.e., 4000-\AA\ break vs. emission-lines).  However, we cannot rule out the possibility that the radio emission may have a large or even dominant AGN component, with a L$_R$-mass relation causing the observed correlation.

\item Composite spectra of the MAIN sample show Balmer absorption strong enough to plausibly swallow any H$\alpha$ emission present.  This is supported by some [NII] emission even in the absence of H$\alpha$.  The [OII] emission, however, increases only slightly with D4000 SFR, giving reason to doubt the validity of D4000 (and thus radio emission) as a proxy of star-formation in these galaxies until we understand why the line emission is so highly-suppressed.  

\item The MAIN quiescent sample shows a correlation between radio luminosity and z-band absolute magnitude of the form: log($L_{1.4}$) = (-0.92 $\pm$ 0.02)$M_z$ + (7.6 $\pm$ 0.4).  The SFRs implied by radio emission reach above 10 M$_{\sun}$ yr$^{-1}$, yet lie just below the objects classified as ``star-forming" for a given absolute magnitude.   If we assume the radio emission stems from SF, one possible interpretation is that the strength of the optical emission lines needed to constitute a ``detection" varies with galaxy luminosity, and using our radio-stacking method allows us to probe low levels of star-formation despite SF dilution in the optical.  If we assume the radio emission stems from AGN activity, then both the D4000 and radio-derived SFRs are too high (for separate reasons), and the observed trend implies a relation between AGN radio luminosity and galaxy mass.  

\item UV-derived SFRs from GALEX are 2-3 orders of magnitude lower than the SFRs predicted by radio-stacking for the MAIN quiescent galaxy sample.  This indicates possible AGN contamination in the radio emission of the quiescent galaxies, incomplete modeling of the dust attenuation for the UV emission, or both.  The different processes at play obviously need to be better understood before we can choose the most useful SFR diagnostic tool for these optically-quiescent galaxies.

\end{itemize}

In short, we use a novel stacking technique to provide strong evidence that quiescent LRGs harbor AGN.    This result has clear implications for galaxy formation scenarios, providing observational support to the existence of a mechanism for low-level feedback in these galaxies.  The results for the quiescent MAIN sample galaxies are more nebulous, with the discrepancies between the different SF indicators clearly highlighting how much we still have to learn about these persistently unclassifiable galaxies.  Overall, we believe the technique utilized here shows great promise as a tool to investigate further the radio emission of various sub-populations in the vast community of radio-quiet galaxies.

\acknowledgments

The authors wish to thank David Helfand for sparking such lively discourse and being a champion of the championless (views, that is).  They also wish to thank Samir Salim for kindly providing the UV data and taking the time to craft such a careful critique of the paper.  JAH acknowledges support of the GAANN Fellowship from the U.S. Department of Education, as well as a UC Davis Graduate Block Grant Fellowship, and Grant HST-GO-10412.03-A from the Space Telescope.  RHB acknowledge the support of the National Science Foundation under grant AST 00-98355.  RLW acknowledges the support of the Space Telescope Science Institute, which is operated by the Association of Universities for Research in Astronomy under NASA contract NAS5-26555.  The work by WDV and RHB was partly performed under the auspices of the U.S. Department of Energy, National Nuclear Security Administration by the University of California, Lawrence Livermore National Laboratory under contract No. W-7405-Eng-48.

Funding for the Sloan Digital Sky Survey (SDSS) has been provided by the Alfred P. Sloan Foundation, the Participating Institutions, the National Aeronautics and Space Administration, the National Science Foundation, the U.S. Department of Energy, the Japanese Monbukagakusho, and the Max Planck Society. The SDSS Web site is http://www.sdss.org/.

The SDSS is managed by the Astrophysical Research Consortium (ARC) for the Participating Institutions. The Participating Institutions are The University of Chicago, Fermilab, the Institute for Advanced Study, the Japan Participation Group, The Johns Hopkins University, Los Alamos National Laboratory, the Max-Planck-Institute for Astronomy (MPIA), the Max-Planck-Institute for Astrophysics (MPA), New Mexico State University, University of Pittsburgh, Princeton University, the United States Naval Observatory, and the University of Washington.

\end{document}